\documentclass[11pt,a4paper]{article}

\usepackage{subcaption}
\usepackage{jcappub}
\usepackage{appendix}
\usepackage{ mathrsfs }
\usepackage{hyperref}
\usepackage{graphicx}

\usepackage{dsfont}
\usepackage{physics}
\usepackage{wrapfig}
\usepackage{epstopdf}
\usepackage{mathtools}

\usepackage{pifont}
\usepackage{amsmath}
\usepackage{amssymb}
\usepackage{xcolor}
\usepackage[T1]{fontenc}

\usepackage{verbatim}

\newcommand{\be}{\begin{equation}}
\newcommand{\ee}{\end{equation}}
\newcommand{\bea}{\begin{eqnarray}}
\newcommand{\eea}{\end{eqnarray}}

\newcommand{\gapp}{\mathrel{\raise.3ex\hbox{$>$}\mkern-14mu \lower0.6ex\hbox{$\sim$}}}
\newcommand{\lapp}{\mathrel{\raise.3ex\hbox{$<$}\mkern-14mu \lower0.6ex\hbox{$\sim$}}}

\newcommand{\half}{\frac{1}{2}}

\def\bbox{{\,\lower0.9pt\vbox{\hrule \hbox{\vrule height 0.2 cm
\hskip 0.2 cm \vrule  height 0.2 cm}\hrule}\,}}

\title{A classical, non-singular, bouncing universe}

\author[a]{\"{O}zen\c{c} G\"{u}ng\"{o}r,} 
\author[a]{and Glenn D. Starkman}
\affiliation[a]{ISO/CERCA/Department of Physics, Case Western Reserve University, Cleveland, OH 44106-7079}

\emailAdd{oxg34@case.edu} 
\emailAdd{gds6@case.edu}

\abstract{
We present a model for a classical, non-singular bouncing cosmology without violation of the null energy condition (NEC).
The field content is General Relativity plus a real scalar field with a canonical kinetic term and only renormalizable, polynomial-type self-interactions for the scalar field in the Jordan frame. 
The universe begins vacuum-energy dominated and is contracting at $t=-\infty$.
We consider a closed universe with a positive spatial curvature, which is responsible for the universe bouncing without any NEC violation.
An $R\phi^2$ coupling between the Ricci scalar and the scalar field drives the scalar field from the initial false vacuum to the true vacuum during the bounce.
The model  is sub-Planckian throughout its evolution and every dimensionful parameter is below the effective-field-theory scale $M_P$, so we expect no ghost-type or tachyonic instabilities. 
This model solves the horizon problem and extends co-moving particle geodesics to past infinity, resulting in a geodesically complete universe without singularities. 
We solve the Friedman equations and the scalar-field equation of motion numerically, and analytically under certain approximations.
}


\notoc
\begin{document}
\maketitle
\flushbottom

\section{Introduction}
\label{section:introduction}
The Inflationary Big-Bang model, our standard model of cosmology, has had notable and numerous observational successes over the last several decades.
Nevertheless,
there has long been considerable interest in alternatives to its earliest universe aspects, specifically in models where the initial Big-Bang singularity is replaced by a big-bounce or by a sequence of bounces. 
Ekpyrotic cosmology is one such longstanding alternative involving branes and extra dimensions \cite{Khoury:2001wf,Khoury:2001bz,Lehners:2007nb,Turok:2004gb}. (See \cite{BounceRevBrandenberger} for a review of progress in this field and many relevant references.) 

Inflation was invented \cite{Guth:1980zm,Starobinsky:1980te} to solve the horizon, flatness and other classic ``problems'' of Big-Bang cosmology. 
Nevertheless inherent shortcomings persist.
Among these is that in Big-Bang cosmology spacetime is not geodesically complete --- the universe inevitably \cite{Hawking:1969sw} originates in a spacelike singularity where curvatures exceed the Planck value, necessitating appeal to an as-yet-unknown ultra-violet-complete theory of quantum gravity.
This has led some to differentiate between the evidence supporting our post-inflationary post-reheating model of cosmology (nucleosynthesis, recombination, growth and evolution of structure), and the far more speculative state of our knowledge preceding that. 
It has also led to many attempts to improve on, or at least develop alternatives to, our theoretical framework for those earliest epochs --
the Hartle-Hawking no-boundary proposal \cite{PhysRevD.28.2960}, 
pre-Big-Bang cosmology \cite{Gasperini:1992em},
ekpyrotic cosmology \cite{Khoury:2001wf} and cyclic cosmology \cite{Steinhardt:2001st}.

A recent alternative to the Inflationary Big-Bang scenario is non-singular bouncing cosmologies \cite{Ijjas_2017,Ijjas:2016tpn}. 
Such cosmologies generically provide geodesic completeness \cite{SteinhardIjjasSimple}, while addressing some of the problems inflation was constructed to solve, such as the horizon problem \cite{SteinhardIjjasSimple}. 
Most models of non-singular cosmologies require a violation of the null energy condition (NEC) \cite{RubakovNEC}. 
This is typically realized with a non-standard kinetic term for a scalar field responsible for the bounce \cite{Ijjas_2017} (and references therein). 
NEC violation and non-standard kinetic terms are not necessarily problematic; however not requiring them would be preferable if possible. Other possibilities for realizing non-singular bounces are models with kinetic braiding \cite{Easson_2011,Dobre_2018}, DHOST models \cite{Ilyas:2020qja}, and theories with torsion \cite{Cubero:2019lxw,Poplawski:2011jz,Unger:2018oqo}.

Recent reanalyses \cite{DiValentino:2019qzk} of Planck data \cite{Planckdata} suggest that we might be living in a universe with a small but positive curvature. The analysis of ACT data alone (\cite{Aiola:2020azj}, see Table 5), on the other hand, are consistent with a flat universe; with the combination ACT+Planck favoring postive curvature at approximately half the significance \cite{Aiola:2020azj}, while ACT+WMAP remains consistent with zero spatial curvature.
When $\kappa > 0$, NEC need not be violated for a cosmological bounce \cite{RubakovNEC}. 
In a scenario with $\Lambda > 0$ and $\kappa>0$, an empty universe will bounce since $\rho_\kappa \sim a^{-2}$ and $\rho_{\Lambda} \sim a^0.$ 
This, however, results in a universe that is symmetric around $t=t_b$ with an equation of state parameter $w=0$. 
Without a field energy density, there is also no mechanism for reheating, and thus no explanation for the current matter content of the universe.

In this paper, we present a model for a clasical non-singular bouncing cosmology without a NEC violation.
This is achieved by having a scalar field with a canonical kinetic term coupled to the Ricci scalar $R$. 
This coupling facilitates the bounce, which occurs when the field-curvature coupling dominates the scale-factor kinematics.
The coupling also prevents the universe from bouncing repeatedly by "locking" the field after the bounce. 

The model we present has only standard kinetic terms and polynomial-type, renormalizable potential terms in the scalar sector in the Jordan frame, and we therefore expect it to be free of tachyonic and ghost-type instabilities.
Dimensionful quantities, such as but not limited to the Hubble constant $H$, the scalar-field amplitude $\phi$, and the Ricci scalar $R$, remain sub-Planckian throughout cosmic history, and the effective-field-theory (EFT) scale remains $M_P$. 
The model has only 2 scales, a mass scale $m$ and the Planck scale $M_P$, and all dimensionless parameters are $\mathcal{O}(1)$ numbers.

As with most non-singular bouncing cosmologies, the model presented here avoids the cosmic singularity problem and is geodesically complete. 
We also provide a solution to the horizon problem without a need for inflation as the current patch comes into causal contact as the universe contracts before the bounce. 
To explain how bouncing cosmologies and the model discussed in this paper solves the horizon problem we will explicitly calculate the horizon size. 

The model exhibits eternal inflation after the bounce, although we anticipate straightforward modifications to permit an end to inflation and reheating. 
As implemented, the model does requires some of this post-bounce inflation in order for the curvature scale today (when the mean photon temperature is $2.7$K) to exceed the current lower limit.

We present a semi-analytic description of the phases in the evolution of the Universe from a large but finite size, through its contraction, then a bounce, followed by a phase of inflationary expansion. 
As discussed below, the evolution can be broken up into epochs during each of which  $\epsilon$, a quantity related to the equation-of-state parameter $w$ of the scalar field, remains approximately constant and can be used to characterize the behaviour of the scale factor $a(t)$, the Hubble parameter $H(t)$, and its time derivative $\dot{H}(t)$. 
We also present full numerical solutions to the Friedmann and the field equations, and show that they agree well with the approximate analytic solutions.

\smallskip
\smallskip
\section{The model} 
\label{section:model}
We assume an isotropic and homogeneous background described by an FLRW metric
\begin{equation}
\label{metric}
\dd s^2 = -\dd t^2 + a^2(t)(\frac{\dd r^2}{1-\kappa r^2} + r^2 \dd \Omega^2),
\end{equation}
where $\kappa>0$ describes a positively curved universe. 
Recent Planck observations (i.e. without inclusion of BAO) favor a slightly positively curved universe with $\Omega_{K,0} \simeq -0.04\pm0.03$ at the $95\%$ confidence level \cite{Planckdata} (and see especially tables 15.1-15.3 of \cite{wiki:2018CosmologicalParameters_MCchains}), 
as emphasized by \cite{DiValentino:2019qzk}.
However, the current magnitude of $\Omega_{K,0}$ is not essential to our model.

We consider a real scalar field coupled to General Relativity.   
The Jordan-frame action is
\begin{equation}
\label{action}
S=\int \sqrt{-g}\dd^4 x \left(\half M_P^2~R -\frac{1}{2}g^{\mu \nu}\nabla_{\mu}\phi \nabla_{\nu}\phi - \frac{\alpha}{2}R\phi^2 - \bar{V}(\phi) \right)\,.
\end{equation}
Unless otherwise stated, we work exclusively in the Jordan frame.
We see that the scalar field has a canonical kinetic term and a simple coupling to the scalar curvature $R$. 
As discussed in the introduction, this coupling plays a central role in the cosmic evolution. 

The scalar-field potential
\begin{equation}
\label{scalarpotential} 
\bar{V}(\phi) = V_0 + \frac{m^2}{2}\phi^2 + \frac{\beta}{3}\phi^3 +\frac{\lambda}{4}\phi^4\, 
\end{equation}
is renormalizable.
$\bar{V}(\phi)$ has a minimum at $\phi_1=0$ (for $m^2>0$).
For appropriate values of $m^2$, $\beta<0$, and $\lambda$
(i.e. $\beta<-\sqrt{4\lambda m^2}\equiv\beta_{max}$), 
$\bar{V}(\phi)$ also has a minimum at
\begin{equation}
	\label{eq:phi0}
	\phi_0 = 
	\frac{-\beta + \sqrt{\beta^2-4m^2\lambda}}{2\lambda}\,,
\end{equation}
and a maximum at
\begin{equation}
	\label{eq:phi2}
	\phi_2 = 
	\frac{-\beta - \sqrt{\beta^2-4m^2\lambda}}{2\lambda}\,.
\end{equation}
If $\beta > -3\sqrt{\frac{\lambda m^2}{2}}\equiv\beta_{min}$,
then $V(\phi_0)>V(\phi_1)$ .
It will be convenient to define
\begin{equation}
\label{scalarpotential} 
V(\phi) = \bar{V}(\phi ) + \frac{\alpha}{2}R\phi^2\,.
\end{equation}
The scalar curvature $R$ acts as a mass term for the scalar field $\phi$, and it can be seen that for $R \gtrsim \abs{\beta/\alpha}$, the potential is a harmonic-like potential with a single global minimum.

We assume throughout that $m\ll M_P$, $\beta/m={\cal O}(1)$, $\lambda={\cal O}(1)$, and $\alpha={\cal O}(1)$
For definiteness, when numerical values are presented or functions are plotted,
we will take 
	$m=10^{-8}M_P$,
	$\alpha = 1/6$,
	$\lambda = 1$, and
	  $\beta = -2.1m$,
	so that $\phi_0 \sim 0.7 m$, and $V(\phi_0)>V(\phi_1)$. 

Although the analysis in this paper is entirely classical, we find it useful to stress that the action \eqref{action}, with the scalar potential $\eqref{scalarpotential}$, has only renormalizable interactions and canonical kinetic terms for the scalar field $\phi$ and the metric $g_{\mu \nu}$.
We therefore expect neither ghost nor tachynonic type instabilities. 
Furthermore the EFT-breaking scale of the theory is $\Lambda \sim M_P$. 
This can be seen by expressing the potential in the Einstein frame
\begin{align}
\label{einsteinpot}
\bar{V}_{E}(\phi) &= V_0\cosh \left(\frac{\phi}{\sqrt{6}M_P}\right) +3 m^2 M_P^2\sinh^2\left(\frac{\phi}{\sqrt{6}M_P}\right)\cosh^2\left(\frac{\phi}{\sqrt{6}M_P}\right) \\ 
&+ 2\sqrt{6}\beta M_P^3\sinh^3\left(\frac{\phi}{\sqrt{6}M_P}\right)\cosh \left(\frac{\phi}{\sqrt{6}M_P}\right) \nonumber\\ &+ 
9\lambda M_P^4\sinh^4 \left(\frac{\phi}{\sqrt{6}M_P}\right) \nonumber
\end{align}
where we have taken $\alpha = 1/6$. 
As it will be shown below that $\phi \ll M_P$,  no extra non-renormalizable terms will be generated. 
We will also show that there is no need for fine-tuning to ensure the suppression of exponentially growing modes, suggesting the absence of tachyonic instabilities.

To solve the Friedman equations, we assume that the classical field $\phi$ behaves as a continuous homogeneous perfect relativistic fluid with stress-energy tensor $T_{\mu \nu} \equiv \frac{-2}{\sqrt{-g}}\fdv{S}{g^{\mu \nu}}$ 
\begin{eqnarray}
\label{stress-energy}
	T_{\mu \nu} &= (1-2\alpha)\nabla_{\mu}\phi \nabla_{\nu}\phi - \frac{1}{2}(1-4\alpha)g_{\mu \nu}g^{\alpha \beta}\nabla_{\alpha}\phi \nabla_{\beta}\phi - g_{\mu \nu} \left(\frac{1}{2}\alpha R \phi^2\ + \bar{V}(\phi)\right) \\ 
	&+\alpha R_{\mu \nu}\phi^2 - 2\alpha \phi \nabla_{\mu}\nabla_{\nu} \phi + 2g_{\mu \nu}\alpha\phi \Box \phi\,. \nonumber
\end{eqnarray}
The energy density $\rho= T_{00}$  and pressure $p$ 
(given by $p g_{ii} = T_{ii}$) are
\begin{align}
	\label{pressureenergy}
	\rho &= \frac{1}{2}\dot{\phi}^2 + \frac{\alpha}{2} R \phi^2 + \bar{V}(\phi) - 3\alpha\frac{\ddot{a}}{a}\phi^2 &\\
	p &= \frac{1}{2}(1-4\alpha)\dot{\phi}^2 -\frac{\alpha}{2}R\phi^2 - \bar{V}(\phi) + \alpha \left(\frac{\ddot{a}}{a}+2\left(\frac{\dot{a}}{a}\right)^2 + 2\frac{\kappa}{a^2}\right)\phi^2 -2\alpha \phi \ddot{\phi} \,,\\
	\intertext{while}
	R &= 6\left(\dot{H} + 2H^2 + \frac{\kappa}{a^2}\right) = 6\left(\frac{\ddot{a}}{a}+\left(\frac{\dot{a}}{a}\right)^2 + \frac{\kappa}{a^2} \right)\,
\end{align}
(where, as is conventional, $H\equiv \dot{a}/a$).

The equation of motion for the scalar field together with the Friedmann equations describe completely the evolution of the universe:
\begin{flalign}
	\label{eq:phieom}
	0 &= \ddot{\phi} + 3 H \dot{\phi} + \pdv{V(\phi)}{\phi}\\
	\label{eq:FriedmannH}
	H^2 &= \frac{1}{3M_P^2}\rho - \frac{\kappa}{a^2}, \\
	\label{eq:FriedmannHdot}
	\dot{H} &= -\frac{1}{2M_P^2}(\rho + p) + \frac{\kappa}{a^2}, \\
	\intertext{and consequently}
	\label{eq:Friedmannaddot}
	\frac{\ddot{a}}{a} &= -\frac{1}{6M_P^2}(\rho + 3p)\,
	.
\end{flalign}
It will be useful below if we express $R$, $H$, $\dot{H}$, and $\ddot{a}/a$ in terms of $\phi$, $\dot{\phi}$, $\ddot{\phi}$ and $a(t)$: 
 \begin{flalign}
 \label{eq:Rphi}
 R &= \frac{\rho-3p}{M_P^2}
 = 
 \frac{4\bar{V}(\phi)-(1-6\alpha)\dot{\phi}^2 
 + 6\alpha \phi \ddot{\phi}}{M_P^2 - \alpha \phi^2}\,,\\
\label{eq:Hsqphi}
H^2 + \frac{\kappa}{a(t)^2} &= \frac{1}{6}\frac{\dot{\phi}^2+2\bar{V}(\phi)}{M_P^2-\alpha\phi^2} \,,\\
\label{eq:Hdotphi}
\dot{H} - \frac{\kappa}{a(t)^2} &=
\frac{(-\half+\alpha)\dot\phi^2 +\alpha\phi\ddot\phi}{M_P^2-\alpha\phi^2}
\,\\
\label{eq:addotbyaphi}
\frac{\ddot{a}}{a} &= -\frac{1}{3}
	\frac{(1-3\alpha)\dot\phi^2 - \bar{V}(\phi) -3\alpha\phi\ddot\phi}{M_P^2-\alpha\phi^2}\,.
 \end{flalign}

The two Friedmann equations, \eqref{eq:FriedmannH} and \eqref{eq:FriedmannHdot}, can be formally solved for $\rho(a)$
\begin{equation}
\label{rhoofa}
\rho (a) = \rho_0 e^{-2\int \epsilon~ \dd \ln a}
\end{equation}
where
\begin{equation}
\label{eos}
\epsilon \equiv \frac{3}{2}(1+\frac{p}{\rho})=\frac{3}{2}(1+w).
\end{equation}
$w$ is the conventional equation-of-state parameter.
For our model, 
\begin{equation}
\label{eosphi}
\epsilon = 3\frac{\left(\frac{1}{2}-\alpha\right)\dot{\phi}^2-\alpha\phi\ddot{\phi}}{\bar{V}(\phi) + \frac{1}{2}\dot{\phi}^2} \,.
\end{equation}

If $\epsilon$ were a constant, \eqref{rhoofa} would straighforwardly give 
\begin{equation}
\label{approxsol}
\frac{\rho}{\rho_i} =  \left(\frac{a}{a_i}\right)^{-2\epsilon}\,,
\end{equation}
where $\rho_i$ is the value of $\rho$ at some fiducial $a=a_i$.
The evolution of the universe might therefore be broken up into epochs in which $\epsilon$ varies very little, followed by large shifts in its value, signalling a different epoch. 

The bounce is a unique moment in time $t=t_b$ when the universe ceases contracting and begins expanding. 
Immediately before the bounce $H<0$, and immediately afterwards $H>0$.
Since this is a non-singular bounce, we insist that the solutions to the equations of motion are continuous, so $H(t_b)=0$.

We fix the arbitrary normalization of the scale factor by choosing $a(t_b)=1$. 
Since  $H(t_b)=0$, 
\begin{equation}
\label{rhob}
\kappa = \frac{\rho_b}{3 M_P^2},
\end{equation}
where $\rho_b\equiv \rho(t_b)$.
$\kappa$ is a free dimensionful parameter. 
It determines the energy density at the bounce.
We choose $\kappa=m^2$, i.e. $\rho_b=3m^2M_P^2$.
In future work we examine a wider range of values  of $\kappa$, however it is clear that we want to keep $\rho_b \ll M^4$.

We can use \eqref{approxsol} and \eqref{rhob} to reexpress  
\begin{equation}
\label{Friedman2}
\begin{split}
H^2 &\simeq \frac{\rho_b}{3M_P^2}a^{-2\epsilon}(1-a^{2(\epsilon-1)}) \\
\dot{H} &\simeq{=} -\frac{\rho_b}{3M_P^2}a^{-2\epsilon}(\epsilon-a^{2(\epsilon-1)})\,. 
\end{split}
\end{equation}
These equations should be understood to hold only when $\epsilon$ is a constant, but are nevertheless useful in describing cosmic evolution during epochs of slowly-varying $\epsilon$. 
The transitions between epochs are obtained by matching the solutions for $a(t)$ and $\phi(t)$ across the boundaries. 

Starting with $a=1$ at $t_b$ and evolving backwards through cosmic time, one can infer several conditions on the available values of $\epsilon$, and on the behaviour of $H^2$ and $\dot{H}$. 
As $H^2 \geq 0$, $\epsilon \leq 1$ before the bounce\footnote{
	This can change after the bounce as other fields are introduced to exit "eternal inflation."
	}. 
For any value of $\epsilon \leq 1$, it can be shown that\footnote{
 	$-\sqrt{H}$ in \eqref{Friedman2} has a minimum for some $a>1$ as $a \rightarrow 1$ from above for $0\leq \epsilon \leq 1$.} 
as $a \rightarrow 1$ from above, $H<0$ decreases towards a minimum, $H_{min}$, where $\dot{H}=0$, before increasing. 
We shall observe this before the bounce, which of course is characterized by $H=0$.
Similarly, after the bounce, as $a$ increases from $1$, $H>0$ will grow to a maximum $H_{max}$.
When $\dot{H}=0$,
\begin{equation}
	\label{eq:aofHmin}
	a(\dot{H}=0) = \epsilon^{1/2(\epsilon -1)}\,,
\end{equation}
so 
\begin{equation}
	\label{Hmin}
	H_{max/min} = \pm\sqrt{\frac{\rho_b}{3 M_P^2}(1-\epsilon)}\epsilon^{-\frac{1}{2(\epsilon -1)}}.
\end{equation}
We will use these markers to label the different epochs of cosmic evolution in the bouncing universe.

In the following sections we first solve the equations of motion approximately and analytically, and discuss the evolution of the universe during different epochs characterized by the equation of state $w=p/\rho$ of the scalar field. 
We will then solve the same equations numerically without any approximations, and show that the resulting solution confirms the conclusions drawn analytically.

\section{Initial Conditions of the Universe}
\label{sec:initialconditions}

We ``begin'' at $t\rightarrow-\infty$ with an infinitely large ($a(t)\rightarrow\infty$) universe, dominated by scalar-field potential energy ${V}(\phi)$. 
We take  $\phi = \phi_0$, and $\dot\phi=0$. 
$\phi_0$ is a minimum of $\bar{V}(\phi)$, not
${V}(\phi)$, but as we show below,
the difference between them, due to $R$, is small.
 
This give us an empty universe dominated by a dark energy density $V(\phi_0)$, and a negative Hubble constant 
\begin{equation}
H(t\rightarrow-\infty) = -\sqrt{\frac{\bar{V}(\phi_0)}{3 (M_P^2-\alpha \phi_0^2)}}\,.
\end{equation} 

 Because $\phi_0$ is a minimum of $\bar{V}(\phi)$,
and $\dot\phi(t\rightarrow-\infty)=0$,
\begin{equation}
	\label{eq:pddotinitial}
	\ddot\phi(t\rightarrow-\infty) 
	= -\frac{\partial V}{\partial\phi}\Bigr\vert_{\phi_0}
	= -\alpha R(t\rightarrow-\infty)\phi_0
\end{equation}
Consequently \eqref{eq:Rphi} gives
\begin{equation}
R(t\rightarrow-\infty) = \frac{4\bar{V}(\phi_0)}{M_P^2-\alpha\phi_0^2}\,,
\end{equation}
and thus
\begin{equation}
	\left\vert
	\frac{6\alpha\phi_0\ddot\phi}
	{4\bar{V}(\phi_0)} 
	\right\vert_{t\rightarrow-\infty} = 
	\frac{6\alpha}{1-\alpha\frac{\phi_0^2}{M_P^2}} 
	\frac{\phi_0^2}{M_P^2}
	= {\cal{O}}\left(\frac{m^2}{M_P^2}\right)
\end{equation}
and 
\begin{equation}
\left\vert
	\frac{\alpha R \phi_0^2}
	{4\bar{V}(\phi_0)} 
	\right\vert_{t\rightarrow-\infty} = 
	\frac{6\alpha}{1-\alpha\frac{\phi_0^2}{M_P^2}} 
	\frac{\phi_0^2}{M_P^2}
	= {\cal{O}}\left(\frac{m^2}{M_P^2}\right)\,.
\end{equation}
The leading-order approximations are:
\begin{eqnarray}
	\label{eq:Rinitial}
	R(t\rightarrow-\infty)&\simeq {4\bar{V}/M_P^2}
	&= {\cal O}(m^4/M_P^2)
	\,,\\
	\label{eq:Hinitial}
	H(t\rightarrow-\infty)&\simeq -\sqrt{\bar{V}/3M_P^2}
	&= {\cal O}(m^2/M_P)
	\,,\\
	\frac{\ddot{a}}{a}(t\rightarrow-\infty)&\simeq 
	\bar{V}/3M_P^2
	&= {\cal O}(m^4/M_P^2)
	\,.
\end{eqnarray}
As promised $\alpha R\phi_0^2$
is initially negligible compared to  
	$m^2\phi_0^2$, $\beta \phi_0^3$, and $\lambda \phi_0^4$.
Thus
\begin{align}
	\bar{V}(\phi)(t\rightarrow-\infty)
		&\simeq V(\phi_0)={\cal O}{(m^4/M_P^2)}\,,&\qquad\qquad\\
	\intertext{but}
	\label{eq:Hdotinitial}
	\dot{H}(t\rightarrow-\infty) 
		&= \frac{\ddot{a}}{a}-H^2 
		\simeq -\epsilon H^2={\cal O}{(m^6/M_P^4)}&\\
	\intertext{is negligible; or equivalently}
	\epsilon(t\rightarrow-\infty)&={\cal O}{(m^2/M_P^2)}\ll1&\,.
\end{align}

We begin, as promised, in an epoch of dark-energy-dominated contraction.

\section{Slow Contraction}
\label{section:slowcontraction}

\begin{figure}[h!]
  \centering
    \includegraphics[width=0.6\linewidth]{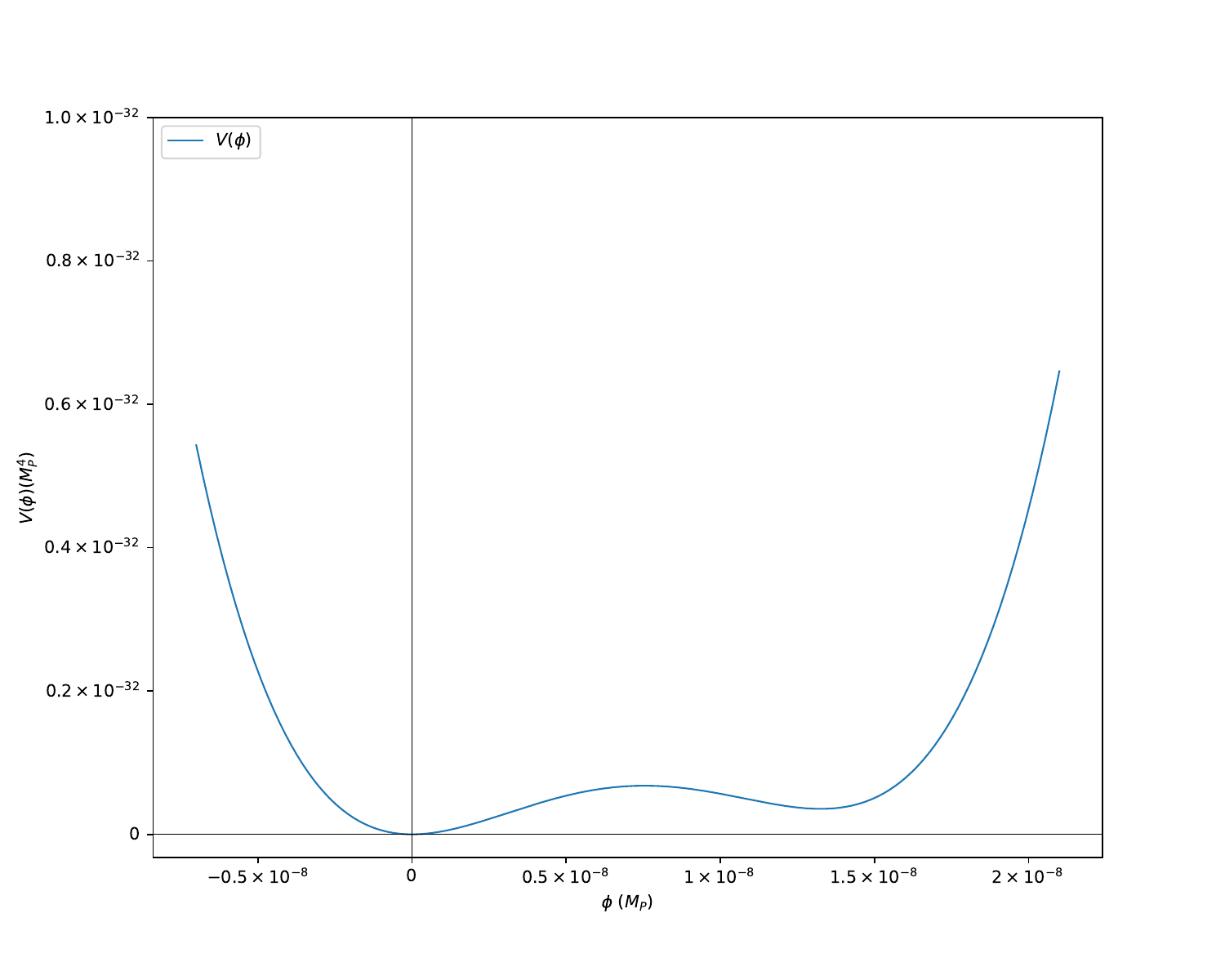}
      \caption{The scalar potential $V(\phi)$ in the case $R\sim0$. }
  \label{fig:potential}
\end{figure}

The first epoch of cosmic evolution is one of slow contraction, with $\epsilon \simeq 0$ , $H \lesssim 0$, and $H$ approximately constant. 
The scale factor can be calculated, assuming a constant $H$,
\begin{equation}
\label{approxscalef}
a(t) = a(t_{early})e^{H_{early}(t-t_{early})} 
\end{equation} 
where $t_{early}$ is a finite value of coordinate time, and $H_{early}$ is given by 
\eqref{eq:Hinitial}.
To examine the behaviour of the field at early times, we expand $V(\phi)$ around $\phi_0(t)$ 
\begin{equation}
\label{adiabaticminima}
\phi_0(t) = \frac{-\beta + \sqrt{\beta^2 - 4\lambda (m^2+\alpha R(t)})}{2\lambda}\,.
\end{equation}
The field will then behave as a harmonic oscillator with \lq negative friction\rq \, of $3H(t)$, following the minimum $\phi_0(t)$ of $V(\phi)$ adiabatically as $R(t)$ increases slowly from its initial negligible value. 
At some early time $t_{early}$ we can take
$\phi(t_{early}) = \phi_0(t_{early})$ and
$\dot{\phi}(t_{early}) = \dot{\phi}_0(t_{early})$,
where
\begin{equation}
\label{incon}
\begin{split}
\dot{\phi}_0(t_{early}) &=  \frac{\alpha \dot{R}(t)}{\beta \lambda\sqrt{1-\frac{4 m^2\lambda}{\beta^2}}}
\end{split}
\end{equation}

To first order in the adiabatic shift 
\begin{equation}
\label{approxphiappr}
	\phi(t) \simeq 
	\phi_0(t_{early}) 
	+ \frac{2 \dot{\phi}_{0}(t_{early})}{3 H \tilde{\omega}}
	e^{-3H(t-t_{early})/2}
	\sin(\frac{3H}{2}(t-t_{early})\tilde{\omega}(t_{early}))
\end{equation}

where
\begin{equation}
\tilde{\omega}(t) \equiv \sqrt{\frac{4\phi_0(t)(\phi_0(t)-\phi_2(t))}{9H^2}-1}.
\end{equation}

At first glance, the form of $\phi(t)$ looks to be one of exponential growth since $H(t < t_{bounce}) < 0$. 
It would then seem to require an infinite fine-tuning of $\phi_0$, since any small perturbation away from $\phi_0$ would grow exponentially as $t_{early}$ is taken to approach $-\infty$. 
This is not the case. 

In the slow contraction phase with the approximation $\dot{H} \simeq 0$
\begin{equation}
\label{Rdot}
\dot{R}(t_{early}) \simeq -12 H\frac{\kappa}{a^2(t_{early})}\,.
\end{equation}
which goes to zero as $t_{early}\to-\infty$.	

The condition for not needing to fine tune in order for any perturbations to not grow exponentially is satisfied if $\frac{2 \dot{\phi}_{0}(t_{early})}{3 H \tilde{\omega}}$ goes to 0 faster than $e^{3Ht_{early}/2}$ as $t_{early}$ is taken to  $-\infty$. 
 Since
 \begin{equation}
 \frac{2 \dot{\phi}_{0}(t_{early})}{3 H \tilde{\omega}} = 
 - \frac{1}{a^2(t_{early})}
 \frac{8\kappa\alpha}
 {\tilde{\omega}
 \beta \lambda\sqrt{1-\frac{4 m^2\lambda}{\beta^2}}}
 \propto e^{-2Ht_{early}} \,,
 \end{equation} 
this no-fine-tuning condition is met.
In other words, as we move $t_{early}$ further and further into the past, the initial condition $\dot\phi(t\to-\infty)=0$ described above is a well-behaved limit that prevents the growing mode from exploding before the bounce. 
This suggests the absence of tachyonic instabilities.

\section{Approaching the bounce}
\label{sec:Approaching}

We have so far taken $H$ to be a constant;
however $\epsilon>0$, and $\dot{H}<0$,
so $H$ slowly decreases as the universe contracts, as can be seen from \eqref{Friedman2}.

To reiterate what was written above, since $\epsilon\leq 1$, $H$ will decrease to some a minimum value $H_{min}$, where $\dot{H}=0$, 
and then increase.  
At $t_b\equiv0$, $H=0$, and the universe ceases contracting and begins expanding.  
After that $H$ increases to $H_{max}$, where again $\dot{H}=0$; and then decreases to a constant -- the beginning of a period of inflation.  
Exiting that inflation requires complicating this model and is not our focus.  

Generically, $\epsilon$ itself will increase from its initial value,
and the scalar-field kinetic terms can become comparable to $V(\phi)$. 
Just how much $\epsilon$ increases depends on the precise values of the various model parameters.
This will determine the duration of the pre-bounce and post-bounce phases -- the period leading up to $H$ reaching its minimum value $H_{min}$, and then growing to $H=0$, and similarly after the bounce.

As discussed above, an increasing scalar curvature $R$ as the bounce nears is crucial to our model. 
A critical time is when $R$ grows large enough that the local minimum $\phi_0(t)$ and the local maximum $\phi_2(t)$ merge to a saddle point. 
This occurs at  
\begin{equation}
R_{crit} 
= \frac{1}{\alpha}\left(\frac{\beta^2}{4\lambda}-m^2 \right)\,,
\end{equation}
freeing the field to oscillate around the global minimum at $\phi = 0$.  
We see that $R_{crit}={\cal O}(m^2)$ is much larger than  $R(t\to-\infty)={\cal O}(m^4/M_P^2)$. 
For $R$ to reach this critical value 
as the universe contracts,
$\dot\phi^2$ must grow to  ${\cal O}(m^2 M_P^2)$.{}
This should therefore happen before, or at least not long after $H \sim H_{min}$, and certainly before the bounce.

For $\alpha = 1/6$ and using \eqref{eq:Rphi} and \eqref{eos}
\begin{flalign}
\label{RHmin}
R &= 6(2-\epsilon)\left(H^2 + \frac{\kappa}{a^2}\right) \\
\intertext{and}
R(H_{min}) &= \frac{4\bar{V}(\phi) + 2\dot{\phi}^2}{M_P^2 - \frac{\phi^2}{6}} - 6\frac{\kappa}{a(H_{min})^2}
\end{flalign}
With $\kappa = m^2$ and $a(H_{min}) = {\cal O}(1)$, for $R$ to reach $R_{crit}$, $\abs{\dot{\phi}^2} \sim {\cal O}(m^2 M_P^2)$. Using \eqref{approxphiappr}, we can estimate the amount of growth needed for the amplitude of $\dot{\phi}^2$ to reach ${\cal O}(m^2 M_P^2)$, i.e. 
\begin{equation}
\label{phidot}
	\abs{\dot{\phi}^2} \sim 144\frac{H^2 \alpha^2 \kappa^2}{\lambda^2 (\beta^2 - 4m^2 \lambda)}e^{-3Ht} 
	= m^2M_P^2\,.
\end{equation}

With $\alpha=1/6$, $\kappa = m^2$, $\beta \sim -2m$ and $\lambda = 1$, this requires $Ht \sim 30$ e-folds in scale-factor growth.
This is the duration of the pre-bounce period --
over the last ~30 e-folds of contraction, $\epsilon$ evolves from being near $0$ to very close to 1, with $\epsilon \rightarrow 1^{-}$ at the bounce.

$\dot\phi^2={\cal O}(m^2M_P^2)$ implies that $\epsilon \sim 1 - {\cal O}(m^2/M_P^2)$. 
This tells us in turn that 
$a(H_{min}) \sim \sqrt{e}$ and thus $R(H_{min}) \sim 3 \kappa$, with ${\cal O}(m^2/M_P^2)$ accuracy.
We have already confined our attention (see above) to $4m^2<\beta^2/\lambda<9m^2$.  
For $\alpha=1/6$, if $\beta^2/\lambda<6 m^2$ then $R(H_{min})>R_{crit}$, and  $\phi$ begins large oscillations before the bounce. 
This suggests that this behavior is typical for natural values of the parameters.
We see this confirmed in the numerical analysis below.

The growth in $R$ is not dramatic from $H=H_{min}$ to the bounce, which is not unexpected since
the universe contracts by a factor of just $\sqrt{e}$. 
For constant $\epsilon$,
we can express $R(a)$ using \eqref{Friedman2} as
\begin{equation}
	\label{scalarR}
	R(a) = 
	6\kappa (2-\epsilon) a^{-2\epsilon}\,.
\end{equation}
Thus between between $H=H_{min}$ and the bounce
$R$ grows by 

\begin{equation}
	R(a=1)/R(H_{min})= 
	\epsilon^{\epsilon/(\epsilon-1)}.
\end{equation}

For $\epsilon \sim 1^-$, this represents an increase of $R$ by a factor of just $\sim e$. 

Once $R \geq R_{crit}$, the potential $V(\phi) = \bar{V}(\phi) + \frac{1}{2}\alpha R \phi^2$ has only a single extremum, a global minimum at $\phi = 0$ around which $\phi$ will start oscillating. From \eqref{eq:Rphi}, at $H = 0$ 
\begin{equation}
\label{atbounce}
	\kappa = \frac{1}{3}\frac{\frac{\dot{\phi}^2}{2}+\bar{V}{(\phi)}}{M_P^2 - \alpha \phi^2}
\end{equation}
in agreement with $\abs{\dot{\phi}^2} = {\cal O}(M^2 m^2)$. Using \eqref{eq:FriedmannH}, \eqref{eq:FriedmannHdot} and \eqref{eos}
\begin{equation}
\label{eosHHdot}
\epsilon = \frac{\frac{\kappa}{a^2}-\dot{H}}{\frac{\kappa}{a^2}+H^2}
\end{equation}
which in turn implies $\dot{H}/\kappa \sim {\cal O}(m^2/M_P^2)$ at or near the bounce.

\section{Focusing on the bounce}
\label{sec:focusonbounce}

Once $R>R_{crit}$, 
the scalar field is free to move from the 
false minimum at $\phi_0(t)$ to the true minimum at $\phi = 0$.
The universe is dominated by scalar field kinetic energy, with $\epsilon \sim 1$.
We show here that $\phi$ is well-behaved through the bounce.

Close to the bounce, we can approximate $H$ as a linear function, 
\begin{equation}
\begin{split}
\label{bounceapprox}
H(t) &\simeq \dot{H}_0 t  \\
a(t) &\simeq e^{\frac{1}{2}\dot{H}_0 t^2}\,.
\end{split}
\end{equation}
Taking $R$ to be its value at the bounce, $R = 6(\dot{H}_0 + \kappa)$, the field equation of motion is
\begin{equation}
\label{approxeom}
\ddot{\phi} + 3 \dot{H}_0 t \dot{\phi} + \left(m^2 + 6\left(\dot{H}_0 + \kappa \right)\right)\phi + \beta \phi^2 + \lambda \phi^3 = 0.
\end{equation}
Expanding to 2nd order in perturbations around a background $\phi = \phi_b + \Delta$
\begin{equation}
\label{approx2}
	\ddot{\Delta} + 3\dot{H}_0 t \dot{\Delta} + \left(m^2 + 6\left(\dot{H}_0 + \kappa \right)\right)(\phi_b + \Delta) + \beta (\phi_b^2 + 2\phi_b \Delta) + \lambda (\phi_b + 3\phi_b \Delta) = 0\,.
\end{equation}
This can be dramatically simplified by definining
\begin{equation}
\label{perturbationeom}
	\tilde{\Delta} \equiv \Delta +  \frac{\eval{\pdv{{V}(\phi)}{\phi}}_{\phi_b}}{\eval{\pdv[2]{{V}(\phi)}{\phi}}_{\phi_b}},
\end{equation}
which satisfies
\begin{equation}
\label{eq:Deltatildeeom}
\ddot{\tilde{\Delta}} + 3\dot{H}_0 t \dot{\tilde{\Delta}} + \tilde{\Delta}\eval{\pdv[2]{{V}(\phi)}{\phi}}_{\phi_b} = 0\,.
\end{equation}

\begin{figure}[h!]
  \centering
    \includegraphics[width=0.6\linewidth]{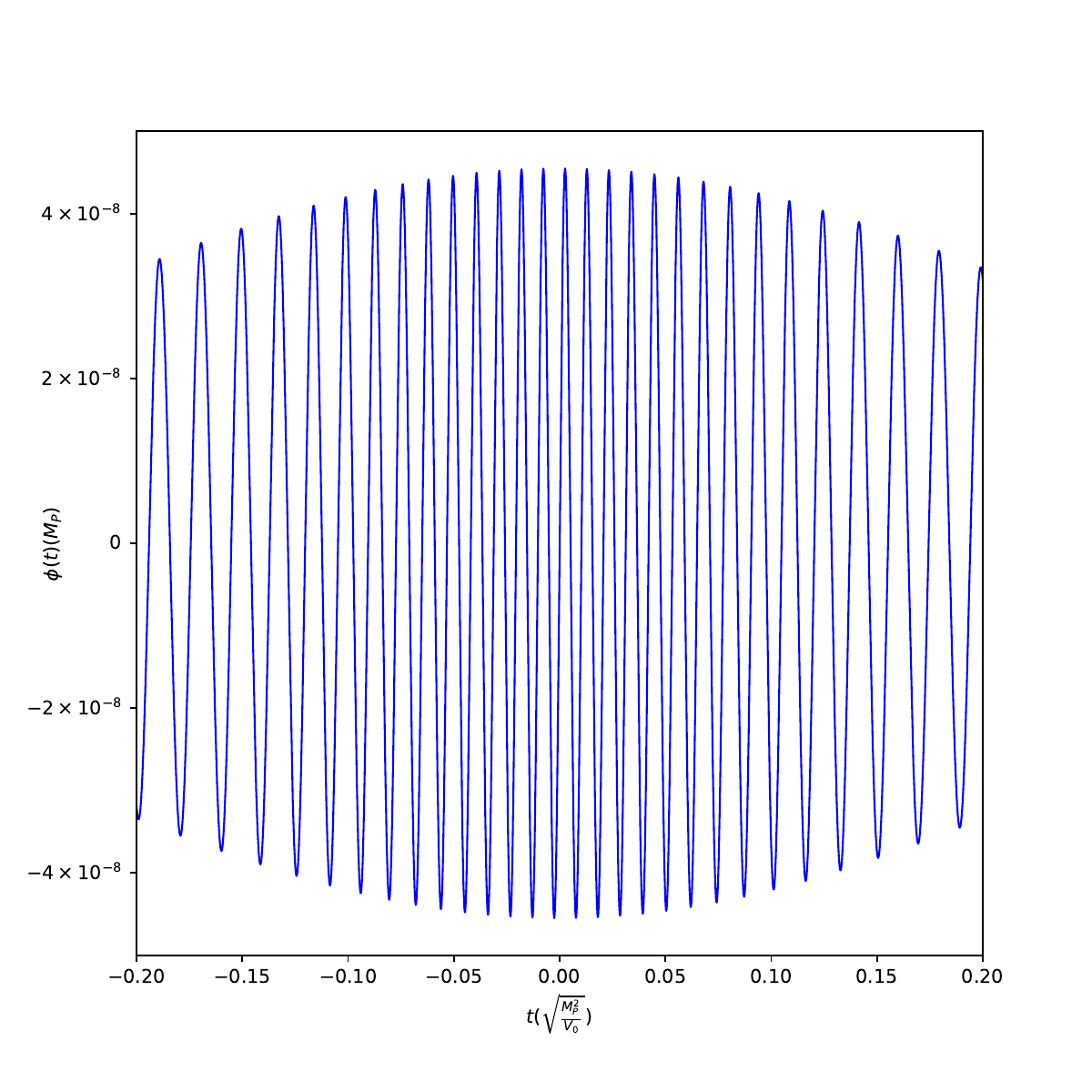}
      \caption{The behavior of the scalar field $\phi(t)$ near the bounce according to \eqref{deltasol}.
       The range of $t$ is chosen such that $\dot{H}_0t^2 \ll 1$ with  $\dot{H}_0 \simeq \kappa(1-\epsilon)\sim 10^{-32}M_P^2$ as obtained in the previous section.}
  \label{fig:approxphi}
\end{figure}

Equation \eqref{eq:Deltatildeeom} can be solved analytically 
\begin{equation}
\label{deltasol}
	\tilde{\Delta}(t) = C \left( e^{-\frac{1}{2}\gamma t^2}H_{\frac{\delta}{\gamma}-1}\left(\sqrt{\frac{\gamma}{2}t}\right) - \frac{\sqrt{\pi}2^{\frac{\delta}{\gamma}-1}}{\Gamma\left(1-\frac{\delta}{2\gamma}\right)}{}_1 F_1\left(\frac{\delta}{2\gamma},\frac{1}{2},\frac{\gamma}{2}t^2\right)\right)
\end{equation}
where $C$ is an overall constant of mass dimension $1$, to be determined later, $\gamma = 3\dot{H}_0$, $\delta = \eval{\pdv[2]{V(\phi)}{\phi}}_{\phi_b}$.
$H_{n}(z)$ are Hermite polynomials and ${}_1F_1(a,b,z)$ is a cofluent hypergeometric function
\begin{equation}
\label{cofluenthypergeometric}
{}_1F_1(a,b,z) \equiv	 \sum_{k=0}^{\infty}\frac{(a)_k}{(b)_k}\frac{z^k}{k!}\,.
\end{equation}
and $(a)_k$ is the Pochammer symbol $(a)_k = \Gamma(a+k)\Gamma(a)$. 
Up to 1st order in $t$, 
\begin{equation}
\label{fieldapprox}
\phi(t) = \phi_b -C\frac{\bar{V}^{\prime}(\phi_b)}{\bar{V}^{\prime\prime}(\phi_b)}\frac{\delta - \gamma}{\sqrt{\gamma}}\frac{\Gamma\left(1-\frac{\delta}{2\gamma}\right)}{\Gamma\left(\frac{3}{2}-\frac{\delta}{2\gamma}\right)} + \tilde{\Delta}(t).
\end{equation}
To determine the value of $C$ and $\dot{H}_0$ we turn our attention to the Friedman equations. Since we know $H = 0$ at the bounce thus $\rho = 3 M^2 \kappa$ and from the previous section we know $\dot{\phi}\sim {\cal O}(M m)$ we can plug in the solution obtained in \eqref{fieldapprox} and \eqref{deltasol} into the expressions in \eqref{stress-energy} to solve for $\dot{H}_0$ and $C$. 
The expressions will not be included here, but the behaviour of the field immediately before and after the bounce is plotted in fig. \ref{fig:approxphi}. 
As discussed above, this solution is valid when $\dot{H}$ can be treated as a constant, i.e. when $\dot{H}t^2 \ll 1$. 
This solution is in agreement with the full numerical solution obtained by solving \eqref{eq:phieom},\eqref{eq:FriedmannH}, and \eqref{eq:FriedmannHdot}, showing that $\phi$ is well-behaved through the bounce.

\section{After the Bounce}
\label{sec:After}

During and slightly after the bounce, with $R \gtrsim R_{crit}$, the full potential $V(\phi)$ has a global minimum at $\phi = 0$. 
After the bounce, a positive $H$ means the oscillations in the field are damped.  
One can expand the potential around $\phi = 0$, and, if $V_0\gg\dot\phi^2$, obtain a solution similar in form to \eqref{approxphiappr} in behaviour where $H$ is now replaced with the constant late-time value $H_{late} = +\sqrt{V_0/3M^2}$:
\begin{equation}
\label{eq:phiapproxab}
	\phi(t) \simeq \frac{2 \dot{\phi}_0}{3 H_{late} \omega^{\prime}}e^{-\frac{3}{2}H_{late} t}\sin(\frac{3 H_{late} \omega^{\prime}}{2} t)\,.
\end{equation}
Here
\begin{equation}
\omega^{\prime} = \sqrt{\frac{4 \phi_0 \phi_2 \lambda}{9 H_{late}^2}-1}\,,
\end{equation}
and for $\phi_0,\phi_2 \sim m$,  $\omega^{\prime} \gg 1$. 

In fact, initially after the bounce $\dot\phi^2\gg V_0$, and $H\gg H_{late}$; however, we can use \eqref{eq:phiapproxab} to get an upper limit on the number of e-folds it would take for $\dot{\phi}^2 \ll V_0$ after the bounce, using the same line of reasoning as in section \ref{sec:Approaching}. 
Since $\omega^{'} \gg 1$, the dominant term in the amplitude of $\dot{\phi}^2$ is
\begin{equation}
\label{phidotab}
\abs{\dot{\phi}^2} = \dot{\phi}_0^2 e^{-3 H_{late} t}.
\end{equation}
From \eqref{eq:Rphi}, $\dot{\phi}_0^2 \simeq 6 m^2 M_P^2$ at the bounce,
and so, with $\dot\phi^2$ falling exponentially, $\abs{\dot{\phi}^2} \sim V_0$ after $N \simeq 10$ e-folds. 
In fact, since initiallly $H\gg H_{late}$, it will take less than $10$ e-folds.

By $N \simeq 15$, $\abs{\dot{\phi}^2} \ll V_0$, 
so $\epsilon \sim 0$, we expect that the universe is dark energy dominated.
The current simple model is then left eternally inflating with $H = H_{late}$.
A graceful exit and reheating requires additional complications, which we reserve to future work. 

After the bounce, as $\dot{\phi}$ decays exponentially, $R$ decreases as well. 
We can only use \eqref{eq:phiapproxab} once $\dot\phi^2\ll V_0$, but thereafter from \eqref{eq:Rphi}, 
\begin{equation}
\label{Rab}
R \simeq \frac{4 V_0 - e^{-3Ht}\dot{\phi}_0^2}{M_P^2}
\end{equation}
and $R \rightarrow {\cal O}(m^4/M_P^2)$ after 10 e-folds.

As discussed further below, the post-bounce inflationary period must last at least  $N\sim 25$ e-folds in order for the universe to satisfy current constraints on the curvature.

\begin{figure}[h!]
  \centering
  	\includegraphics[width=0.99\linewidth]{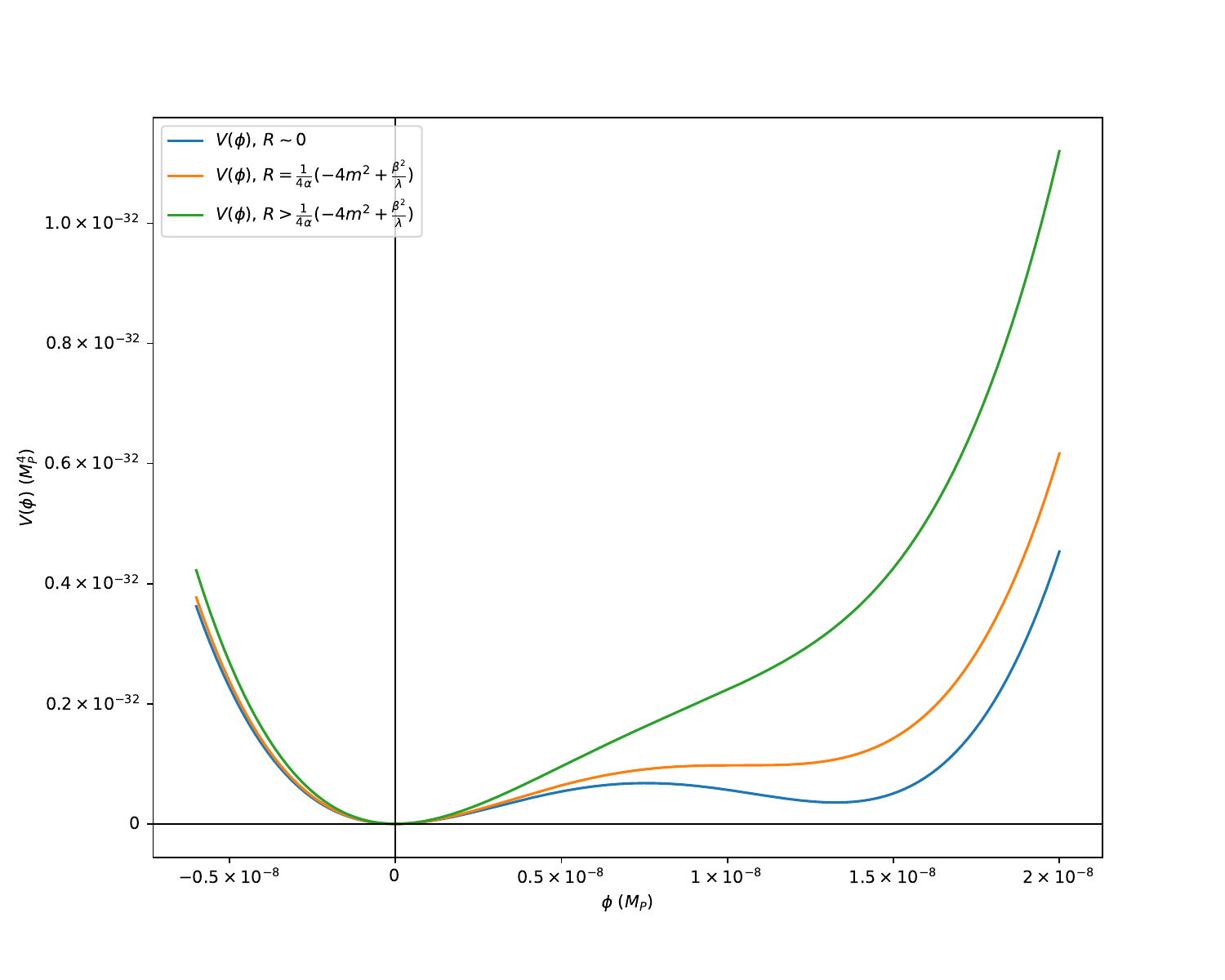}
  	\caption{The scalar potential $V(\phi)$ at selected values of the scalar curvature $R$. The blue curve is the initial scalar potential when $R \sim 0$ and has two clear minima; the orange is for $R = \frac{1}{4\alpha}\left(-4m^2 + \frac{\beta^2}{\lambda}\right)$, when $V(\phi)$ has a saddle point instead of a second minimum; the green has $R > 2\frac{1}{4\alpha}\left(-4m^2 + \frac{\beta^2}{\lambda}\right)$, and $V(\phi)$ has just one minimum.}
  	\label{fig:pot2}
 \end{figure}

\section{Numerical solution}
We now turn our attention to numerically solving the system of equations in \eqref{eq:phieom},  \eqref{eq:FriedmannHdot}, and \eqref{eq:Friedmannaddot}. 
The dimensionless free parameters are taken to be 
$m/M_P = 10^{-8}$, $\beta/m = -2.1 $, $\lambda = 1$, $\alpha = 1/6$, $\kappa/m^2 = 1.0 $, $V_0/m^4 = 0.0001$. A wider exploration of the phenomenology of the model parameter space will be reported on in futue work.

Numerically, we start with the field at the false minimum at $\phi = \phi_0$ and some arbitrary initial value of $a=a_i$. 
We integrate forwards in time to and then through the bounce. 
We then re-scale $a(t)$ such that $a(t_b) = 1$, and shift the origin of time such that $t_b = 0$.

As the scale factor evolves, so too does the scalar curvature $R$.  
The coupling between the scalar field $\phi$ and  $R$, drives a change in the scalar potential $V(\phi)$.
When $t \sim - \sqrt{M_P^2/V_0}$, 
$R \sim R_{crit}$, and $V(\phi)$ develops a saddle point.
The scalar curvature term in $V(\phi)$ soon dominates the shape of the potential, and the potential resembles a harmonic well with a minimum at $\phi=0$.
After the bounce $R$ becomes small again and no longer dominates $V(\phi)$.
The shape of the potential in these epochs is plotted in fig. \ref{fig:pot2}.

The numerical solution for $\phi(t)$ shows it to behave at early times like a forced oscillator adiabatically following the local minimum, in agreement with the approximations discussed above.
Near and during the bounce, as $R\geq R_{crit}$, $\phi(t)$ oscillates around the global minimum of  $V(\phi)$ at $\phi=0$. 
Meanwhile $\epsilon$ evolves towards unity signalling kinetic energy domination.
The Hubble parameter crosses zero and the universe bounces.
As discussed above, there is a periods of  $\sim 35$ e-folds before the bounce and $\sim 10$ after the bounce during which $\dot{\phi}^2$ is comparable to or dominates $V(\phi)$, and before and after which the universe is dark energy dominated. 

Post-bounce, $\phi(t)$ is a damped oscillator around the global minimum at $\phi=0$. 
After a phase of rapid expansion lasting $\sim 10$ e-folds, the universe inflates eternally with $H^2 = \frac{V_0}{3M_P^2}$. 

\begin{figure}[h!]
  \centering
    \includegraphics[width=0.8\linewidth]{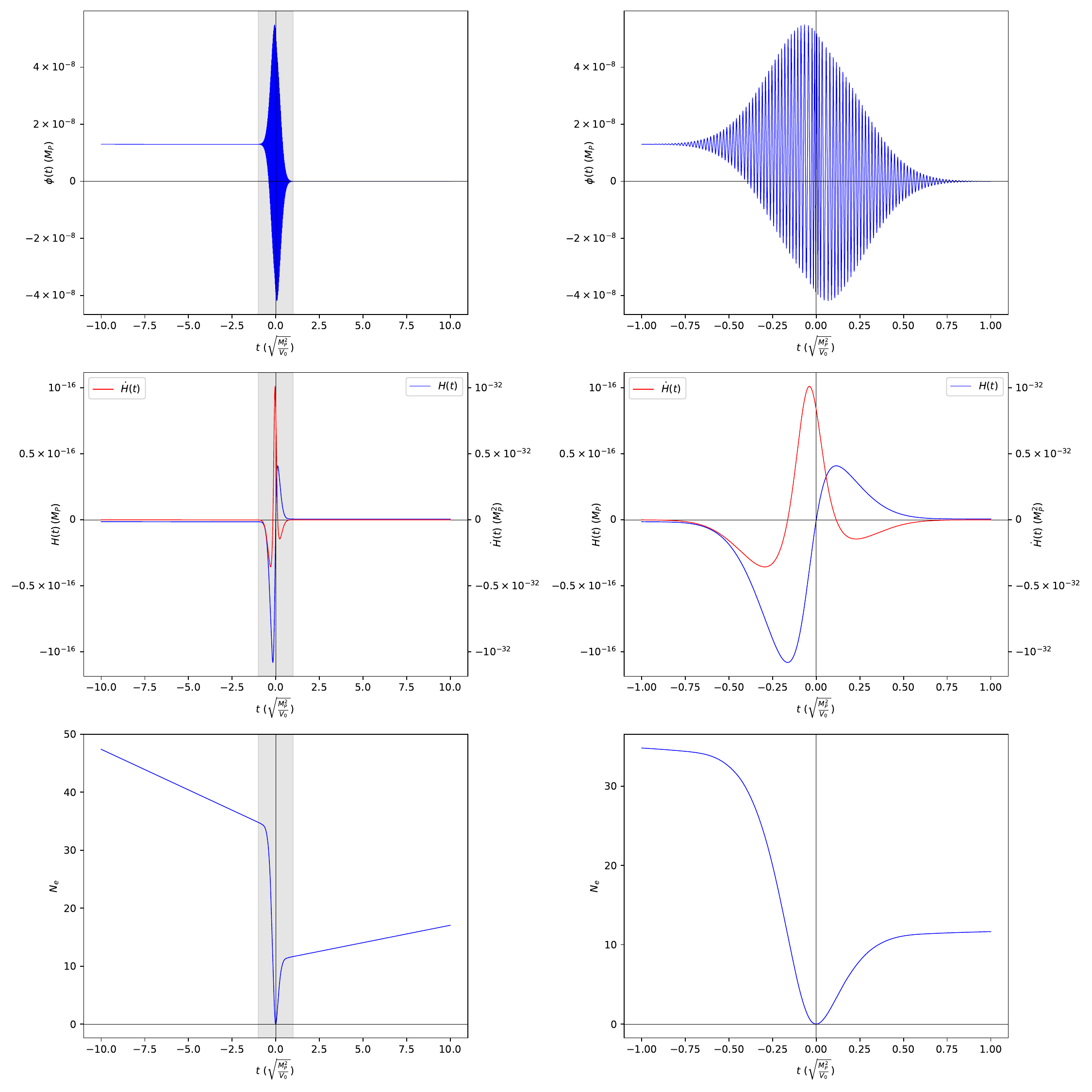}

  \caption{The behaviors of the field $\phi(t)$ and cosmological parameters of interest $N_e, H(t), \dot{H}(t).$. The plots on the left show overall behavior and the plots on the right are zoomed in versions of the shaded grey areas concentrating on the bounce. 
  $t$ is given in units of $\sqrt{M_P^2/V_0}$ to naturally encompass the bounce phase in the range $-1\leq t \leq 1$, and to naturally include $\sim 100$ oscillations of $\phi$ with  period of $\sim \frac{1}{m^2 + (\sqrt{H^2 + \dot{H}})}$.}
  \label{fig:scalefactorandH}
\end{figure}

The full numerical solution is in agreement with the approximate analytic solutions obtained in section 4-6. 
As can be seen in fig. \ref{fig:scalefactorandH}, it takes $N\sim 30$ e-folds to transition from a dark-energy-dominated universe where $\epsilon \sim 0$ to one with $\epsilon \lesssim 1$. 
In fig. \ref{fig:scalefactorandH} one  can clearly identify different cosmic epochs: $\epsilon \sim 0$ in early times; followed by a transition to $\epsilon \lesssim 1$ lasting for $\sim 40$ e-folds; after which the universe is stuck in eternal inflation with once again $\epsilon \sim 0$. 
With the parameters chosen above, $H_{late} \sim 10^{-18} M_P$.

\section{The Horizon Problem and other Cosmological Problems}

As stated in the introduction, non-singular bouncing cosmologies can naturally solve the horizon problem -- the fact that the universe appears to be homogeneous on length scales greater than naively should be  expected, namely no more than twice the apparent horizon.
This is the distance that light would have been able to travel from the big bant to the time the relevant signal was emitted -- i.e. the particle horizon -- but in a universe with only the currently observed contents -- matter, radiation, and dark energy. 

This problem is most clear in the extreme isotropy of the cosmic microwave background (CMB) over the whole sky despite  that in a matter-and-radiation-dominated big-bang cosmology, the particle horizon size at last scattering is only $\sim1^\circ$.

Up to a factor of order unity, the (non-inflationary) big-bang particle-horizon at time  $t$ is the Hubble scale
\begin{equation}
\label{horizon}
r_H = \frac{1}{\abs{H(t)}}.
\end{equation}
This is to be contrasted with the radius of curvature of the constant curvature hypersurface,
$r_c(t) \equiv a(t)/\sqrt{\kappa}$, so that $r_0 = 1/\sqrt{\kappa}$.
Since the curvature is positive, this hypersurface is compact and of finite extent.
$r_c(t)$ is greater than or equal to the ``size of the universe.''
Since observationally $\Omega_k\lesssim 0.04$,
$r_0\gapp 15 h^{-1}$Gpc, which is  greater than the radius of the observable universe, i.e. the last scattering surface of the CMB.  
Therefore, if the actual particle horizon at any time $t$ is greater than $r_c(t)$, we have solved the horizon problem.

This happens naturally in the model we describe. After $H=H_{min}$, $\epsilon$ is approximately constant and the ratio of the Hubble scale to the curvature scale is 
\begin{equation}
\label{horizon2}
\frac{r_H}{r_c} = \frac{a^{\epsilon-1}}{\sqrt{1-a^{2(\epsilon -1)}}}
\end{equation}
which is greater than 1 for any $\epsilon > 0$. 
As explained above, $\epsilon \lesssim 1$ in between $H_{min}$ and $t_{bounce}$; thus there exists a period  during which the whole universe was within one Hubble scale.

Of course the Hubble scale diverges at the bounce, whereas the particle horizon is finite, so
using \eqref{horizon} as a stand in for the particle horizon is not sufficient.
A more careful analysis involves using the proper integral definiton of the particle horizon 
\begin{equation}
\label{horizonint}
r_p(t) = a(t)\int_{-\infty}^t\frac{\dd t^{\prime}}{a(t^{\prime})}
\end{equation}
This can be shown to be finite at all finite times $t$, including $t=t_b$. 

The appearance of $-\infty$ as the lower limit of the integral for $r_p$ may look unfamiliar but  is crucial in a bouncing cosmology, since geodesics are naturally extended to past timelike infinity.
This is in contrast to the calculation of the horizon size in a Big-Bang cosmology since the earliest time a signal can be generated is the Big Bang singularity.

\begin{figure}[h!]
	\centering
		\includegraphics[width=0.9\linewidth]{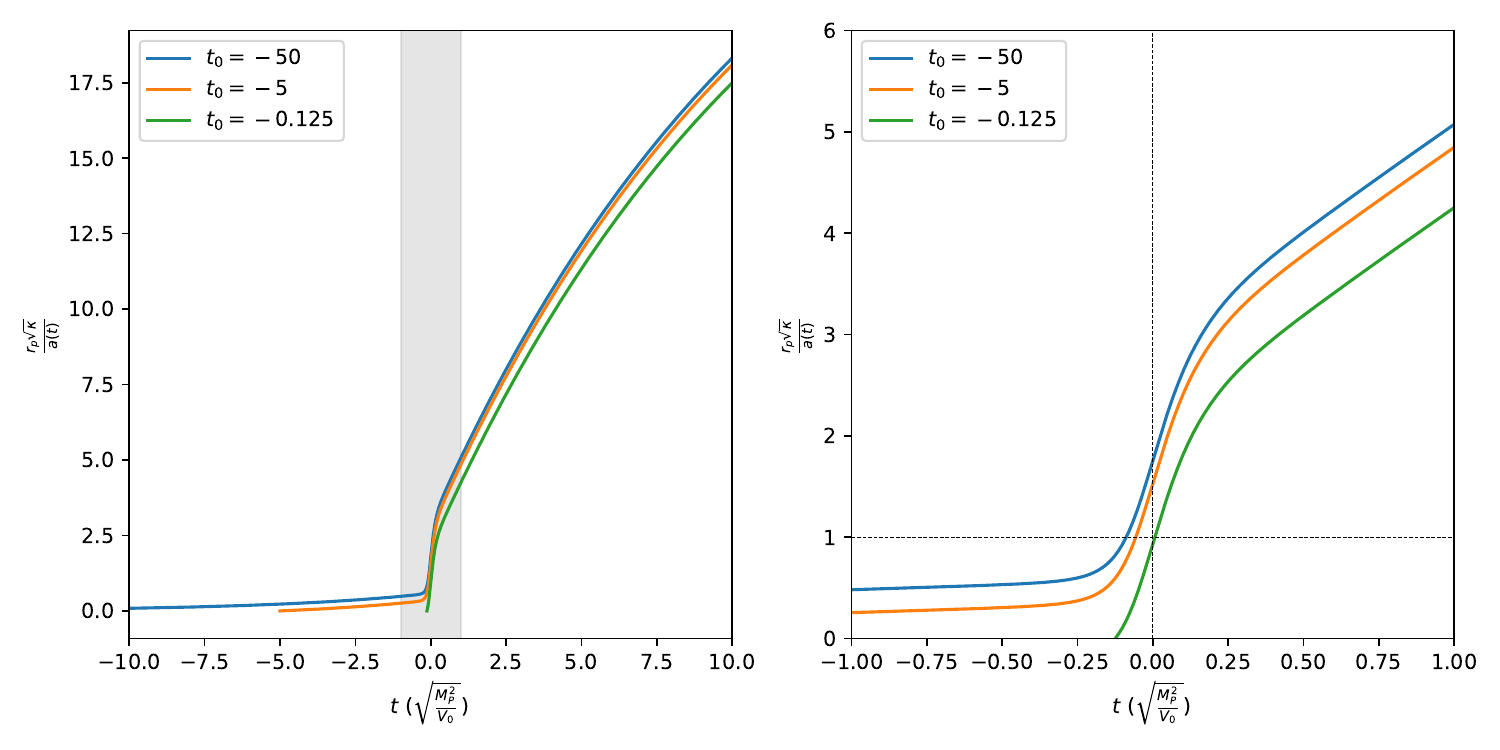}
			\caption{On the left: the ratio of the particle horizon $r_p(t)$ and the radius of curvature of the constant curvature hypersurface $r_c(t) = a(t)/\sqrt{\kappa}$ for the same three signals. At $t=t_b\equiv0$ the signals that have originated at $t_0=-50$, $t_0=-5$ have particle horizons larger than $r_c(t)$ therefore they are in causal contact. The signal that originated at $t=-0.125$ comes into causal contact at the bounce and any signal that originates at $t_0 > -0.125$ would come into causal contact sometime after the bounce. The rapid rise in the ratio is due to the integral in \eqref{horizonint} receiving almost all of its contribution when $a(t)$ is small near the bounce. On the right: the same plot zoomed in, showing the different $r_p(t)/r_c(t)$ ratios near the bounce.}
		\label{fig:horizonsdouble}
\end{figure}

The above defined $r_p(t)$ is the maximum distance a particle could travel if that particle was in existence at the infinite past. 
If a particle was created at a later time $t_0$, its particle horizon would have the integral start from $t=t_0$. 
Particles created at earlier times would have larger particle horizons. 
We illustrate the solution to the horizon problem by calculating the ratio $r_p(t)/r_c(t)$ for 3 different initial times $t_0=-50$,$-5$,$-0.125$ (in our customary units of $\sqrt{M_P^2/V_0}$) in figure \ref{fig:horizonsdouble} using the numerical solution obtained above for $a(t)$.  
The case $t_0=-50$ is functionally indistingishable from $t_0=-\infty$ since most of the contribution to $r_p(t)$ is when $t\sim t_b$ during which $a(t)\sim 1$.
We have chosen $t=-0.125$ because 
only if $t_0\gapp-0.125$ is $r_p(t)/r_c(t)<1$ by $t=0$.

The solution to the horizon problem offered by bouncing cosmologies is thus quite simple -- 
the bounce brought the entire universe within causal contact.
For any point in the universe, starting at $t<-0.125$, its forward light cone encompasses the entire universe. 
Therefore the past light cones of any two points in the universe at the bounce intersect at all times earlier than $t=-0.125$.

The evolution of other cosmological observables such as the cosmic curvature factor $\Omega_{k}$ and the cosmic anisotropy factor $\Omega_{a}$ is of importance as well. The cosmic curvature factor is defined as
\begin{equation}
\label{cosmiccurvature}
\Omega_k \equiv -\frac{\kappa}{a^2}H^{-2}.
\end{equation}
Using the constant-$\epsilon$ approximation \eqref{Friedman2},   
\begin{equation}
\label{cosmiccurvature2}
\Omega_k = -\frac{a^{2\epsilon}}{a^2 - a^{2(\epsilon+1)}}.
\end{equation}.

For $0 < \epsilon \lesssim 1$, $\Omega_k$ increases as the universe approaches the bounce and is divergent at the bounce when $H=0$. After the bounce, for $0 \leq \epsilon \lesssim 1$, $\Omega_k$ decreases. The divergence, however, is not of issue because these quantities are of interest only after the bounce. Current experiments suggest $\frac{\Omega_k}{\Omega_\Lambda} \lesssim 0.1$ and for $\kappa = 10^{-16}M_P^2$ and $V_0 = 10^{-36}M_P^2$ this would require  $N\sim 25$ e-folds of post-bounce inflation. In other words, however this toy model is augmented to permit a graceful exit from inflation, it must happen after at least this many e-folds.

It is important to note that this number, $N\sim 25$, is dependent on model parameters, and it is conceivable that there are regions of parameter space requiring litle or no inflation.  
There would still be a need to exit inflation, and efficiently reheat.

The cosmic anisotropy factor is defined as
\begin{equation}
\label{cosmicanisotropy}
\Omega_a \equiv \frac{\sigma^2}{a^6}H^{-2}
\end{equation}
and again using the constant-$\epsilon$ approximation \eqref{Friedman2}, 
\begin{equation}
\label{cosmicanisotropy2}
\Omega_a = \frac{\sigma^2}{a^6}\frac{1}{1- a^{-2\epsilon}}.
\end{equation}

As the universe approaches the bounce, $\Omega_a$ increases and is divergent at the bounce when $H=0$. The divergence, as in the case of $\Omega_k$ is of no worry as only the post bounce values of these observables are of interest. Nevertheless, as the universe contracts for an infinite amount of time before the bounce, $\Omega_a$ would grow as $a^6$ and since the expansion phase post-bounce lasts for a finite amount of time, there is no natural mechanism to ``erase'' the anisotropies generated during the infinitely long phase of contraction. In this work, we assume $\sigma^2 = 0$ exactly and assume the scalar field $\phi$ and the metric $g_{\mu \nu}$ to be perfectly isotropic. The analysis of anisotropies generated by quantum fluctuations in the scalar field, in the metric, and in Standard Model fields, and their evolution is the subject of future work.

\section{Conclusion}
In this paper we have constructed a model for a non-singular, non-NEC violating cosmological bounce scenario using a standard scalar field to mediate the bounce. 
We achieve this by coupling the scalar field to the scalar curvature $R$. 
This coupling has several key consequences: it forces the scalar field to move from the false minimum towards the true minimum before the bounce; it ensures the field stays in the true minimum after the bounce; and by increasing the frequency of oscillations of the scalar field, it drives the density of state paremeter $\epsilon$ close to $1$. 

Our model has a postively curved compact universe with $\kappa > 0$.  
This allows the bounce to occur without violating the null energy condition.
A closed universe with positive curvature is not ruled out by current observations, and may even be slightly favored, albeit with very small curvature.  
With the limited subset of model parameter space that we have explored, this necessitates some amount (about 25 e-folds) of post-bounce inflation.  
There may  be values of the model parameters for which this is not necessary.

The analysis in this paper is entirely classical but it is informative to discuss the model in an EFT framework. 
Our model has only canonical kinetic terms for the fields and the scalar potential includes only renormalizable terms of the usual polynomial type. 
Canonical kinetic terms suggest that the model is free of ghost instabilities, and the renormalizability of the scalar potential ensures that the EFT breaking scale stays at $M_P$. 
As we have shown in sections \ref{sec:Approaching} and \ref{sec:focusonbounce}, there are no exponentially growing scalar-field modes in danger of exploding before the bounce, and the field is well-behaved during the bounce, suggesting the absence of tachyonic instabilities. 

In our toy model, the post-bounce universe is eternally inflating, so complications would be needed to allow a graceful timely exit from inflation with appropriate reheating.

In future work, 
we will more carefully explore the stability of this model to ghost, tachyonic and especially gradient instabilities.
We will consider in greater detail the generation and evolution of fluctuations that can seed structure in the universe, as well as a concrete model for exiting inflation. 
These must remain subdominant through the infinitely long period of pre-bounce cosmic contraction, then grow to significance post-bounce post-inflation post-reheating. 
As pointed out recently \cite{Agullo:2020fbw}, a bounce in the history of the universe may leave imprints in the non-gaussianities in perturbations which might lead to observable consequences for bouncing cosmological models.

Although many features remain to be explored, it is intriguing that the universe may have begun nearly empty and contracting, and our big-bang-like observable universe is the aftermath of a non-singular NEC-respecting classical bounce.

\acknowledgments
OG and GDS are partially supported by grant DOE-SC0009946 from the US Department of Energy. OG and GDS thank Kurt Hinterbichler and the late Bryan W. Lynn  for their helpful discussions. 

\bibliography{bounce-pub.bib}
\bibliographystyle{unsrt}


\end{document}